# Half-cycle attosecond pulse-induced dynamics of space-time photonic crystals

Rostislav Arkhipov[1,2*]

[1]St. Petersburg State University, St. Petersburg, 199034 Russia
[2]Ioffe Institute, 194021 St. Petersburg, Russia
*arkhipovrostislav@gmail.com

Rapidly changing the refractive index of a medium in space and time (space-time photonic crystal, STPC) has been a challenging task. Such a rapid change can be achieved by carrier-wave Rabi flopping. We show that it can be realized when a train of half-cycle light pulses collide in a simple three-level atomic medium. We found the formation of Bragg microcavities and demonstrated their ultrafast control in this case. The proposed approach is a convenient way to control medium properties on the sub-cycle time scale. It is a demonstration of the formation of STPCs and their ultrafast control in a simple atomic medium, which has been a difficult task until now.

## 1. Introduction

Extremely short pulses of light became a remarkable tool for studying the dynamics of electrons in matter [1-3]. The award of the Nobel Prize in Physics in 2023 was in recognition of the advances in this field [4]. A further reduction in pulse duration in a particular spectral region can be achieved by removing all the half-waves from the conventional multi-cycle pulse and retaining only a single half-wave. Such a pulse is referred to as a half-cycle pulse or a unipolar pulse [5-11]. Its important characteristic is the electrical pulse area $S_E$. This is defined as the integral of the electrical field strength $E$ over time at a given point in space $r$ [5-7]:

$$S_E = \int E(r,t)dt, \qquad (1)$$

This quantity is important in physics because the electrical area determines how such pulses affect atoms, molecules and nanostructures [12-14]. Due to the unidirectional ultrafast momentum transfer to the electron, such a pulse with a large electric area can serve as an efficient tool for ultrafast control of quantum objects, holographic recording of ultrafast moving objects and other interesting applications, see reviews [5,6,11] and the literature therein.

In parallel, space-time photonic crystals (STPCs), materials whose refractive index varies both with time, and photonic time crystals (PTCs) [15], materials whose refractive index changes abruptly with time, have recently become the subject of active interest in ultrafast science, see reviews [16-18]. When the refractive index changes rapidly in time and space, a number of unusual phenomena can occur. These include time refraction and time reflection of light [19,20]. They can occur on the timescale of a single optical cycle, in femtosecond units. Simultaneously, the recent study of such media has paved the way for investigating many interesting new phenomena [15-17]. Such a material can also serve as a perspective material for the development of a new type of laser source [21].

Despite the great potential of PTCs and STPCs, their experimental realization is challenging. Conventional nonlinear optical mechanisms that allow the refractive index to change are too weak or too slow [16-18,20]. Therefore, some kind of novel materials or other physical mechanisms are needed to realize fast refractive index changes. To date, various types of exotic materials, such as transparent conducting oxides (TCOs), which operate close to their epsilon-near-zero regime, have recently been found to exhibit very large and fast light-induced refractive index changes, see [20,22]. Therefore, the search for another mechanism for ultrafast and dramatic changes in the state of matter in conventionally available media is still crucial.

Carrier wave Rabi flopping (CWRF) is one of the mechanisms that allows rapid state changes in matter [23]. It occurs when a single- or sub-cycle light pulse interacts with an atomic medium [24-25]. This leads to a rapid change in the atomic population at the Rabi frequency. This frequency occurs within the duration of the pulse and is comparable to the transition frequency of the medium. The CWRF phenomenon was predicted in Ref. [23] and has since been observed in semiconductors [26], atoms [27,28], etc.

Half-cycle pulses, as mentioned above, have the shortest possible duration within a given spectral range [7]. They can therefore be used as a powerful tool for to realize ultrafast changes in the state of the medium [13,14]. In fact, such pulses can form gratings of atomic populations via CWRF if they interact coherently with the medium (the pulse duration and the delays between pulses are shorter than the polarization relaxation time $T_2$ of the medium) [29-31]. It is also possible to control these gratings. They can be created and erased, and their spatial period can be multiplied on the time scale of the pulse duration [29,30], see also review [32]. In addition, a dynamic microcavity can be formed when rectangular unipolar pulses collide in a two-level resonant medium [33]. This microcavity can have Bragg-like mirrors at its boundaries [34]. In addition, a grating can be formed by the collision of two counterpropagating few-cycle self-induced transparency solitons in the resonant medium in their overlap region [35].

Therefore, PTCs and STPCs can be practically implemented by CWRF using extremely short pulses. In this paper, we demonstrate a simple method for the generation and control of PTCs and STPCs in a simple three-level atomic medium using pairs of half-cycle attosecond pulses that collide in the medium. We find that in this case the population difference has a constant value in the center of the medium in the region where the pulses overlap. On the sides of this region, a periodic grating of populations is formed. This means that at each resonant transition in the medium, a microcavity with Bragg-like mirrors is created.

This is what makes the case discussed here different from that of a two-level medium [34], where the microcavity is formed by the collision of single-cycle self-induced transparency (SIT) pulses. In this Ref. [34], however, the microcavity formed only near the pulse overlap region. And the resulting Bragg grating had only a few periods, which limited its potential applications. Here we show that by applying a train of half-cycle pulses, it is possible to control the microcavity in an ultrafast manner. The Bragg grating fills the entire medium from both sides of the pulse overlap. We also show that using SIT pulses to form gratings, as previously considered [33-35], is not necessary. The proposed method could be a more convenient way of ultrafast generation and control of SPCs and STPCs in a simple atomic medium.

## 2. The model and problem statement

The analytical calculations of the grating dynamics can be easily performed for the two-level medium in the case of non-overlapping pulses [29,30]. For the multilevel medium, the gratings can be simply calculated by means of the approximate solution of the time-dependent Schrödinger equation in the weak-field regime [31]. Under the assumption that the medium is affected by the pair of half-cycle delta-pulses, $E(t) =$

$S_{E1}\delta(t) + S_{E2}\delta(t - \Delta)$, with electric pulse areas, $S_{E1}$ and $S_{E2}$, with the delay between them being $\Delta$, the population of the k-th bound state is given by the following expression: $w_k = \frac{d_{0k}^2}{\hbar^2}(S_{E1}^2 + S_{E2}^2 + 2S_{E1}S_{E2}\cos\omega_{0k}\Delta)$, $d_{0k}$ is the transition dipole moment, $\omega_{0k}$ is the transition frequency. This formula for population values is very similar to that for the total intensity of two interfering monochromatic light beams. However, in the case of half-cycle pulses, their interference in the conventional sense is not possible. This formula shows that harmonic grating can be formed due to the "interference" of the electrical areas of the pulses [36].

This simple approach, cannot take into account the spatial-temporal dynamics of the medium polarization. In the following, we consider a three-level medium affected by strong half-cycle pulses colliding in the medium. In this case the medium excitation is strong. CWF occurs and the simple analytical solution is unknown. Therefore, we study this problem numerically using the well-known system of Maxwell-Bloch equations, including material equations for non-diagonal and diagonal elements of the density matrix elements, coupled to the wave equation for the electric field. This system is written beyond the slowly varying envelope and rotating wave approximations and is given by [37]

$$\frac{\partial}{\partial t}\rho_{21} = -\rho_{21}/T_{21} - i\omega_{12}\rho_{21} - i\frac{d_{12}}{\hbar}E(\rho_{22} - \rho_{11}) - i\frac{d_{13}}{\hbar}E\rho_{23} + i\frac{d_{23}}{\hbar}E\rho_{31}, \quad (1)$$

$$\frac{\partial}{\partial t}\rho_{32} = -\rho_{32}/T_{32} - i\omega_{32}\rho_{32} - i\frac{d_{23}}{\hbar}E(\rho_{33} - \rho_{22}) - i\frac{d_{12}}{\hbar}E\rho_{31} + i\frac{d_{13}}{\hbar}E\rho_{21}, \quad (2)$$

$$\frac{\partial}{\partial t}\rho_{31} = -\rho_{31}/T_{31} - i\omega_{31}\rho_{31} - i\frac{d_{13}}{\hbar}E(\rho_{33} - \rho_{11}) - i\frac{d_{12}}{\hbar}E\rho_{32} + i\frac{d_{23}}{\hbar}E\rho_{21}, \quad (3)$$

$$\frac{\partial}{\partial t}\rho_{11} = -\frac{\rho_{22}}{T_{22}} + \frac{\rho_{33}}{T_{33}} + i\frac{d_{12}}{\hbar}E(\rho_{21} - \rho_{21}^*) - i\frac{d_{13}}{\hbar}E(\rho_{13} - \rho_{13}^*), \quad (4)$$

$$\frac{\partial}{\partial t}\rho_{22} = -\rho_{22}/T_{22} - i\frac{d_{12}}{\hbar}E(\rho_{21} - \rho_{21}^*) - i\frac{d_{23}}{\hbar}E(\rho_{23} - \rho_{23}^*), \quad (5)$$

$$\frac{\partial}{\partial t}\rho_{33} = -\frac{\rho_{33}}{T_{33}} + i\frac{d_{13}}{\hbar}E(\rho_{13} - \rho_{13}^*) + i\frac{d_{23}}{\hbar}E(\rho_{23} - \rho_{23}^*), \quad (6)$$

$$P(z,t) = 2N_0 d_{12} Re\rho_{12}(z,t) + 2N_0 d_{13} Re\rho_{13}(z,t) + 2N_0 d_{12} Re\rho_{12}(z,t) + 2N_0 d_{23} Re\rho_{32}(z,t). \quad (7)$$

$$\frac{\partial^2 E(z,t)}{\partial z^2} - \frac{1}{c^2}\frac{\partial^2 E(z,t)}{\partial t^2} = \frac{4\pi}{c^2}\frac{\partial^2 P(z,t)}{\partial t^2}. \quad (8)$$

Here $N_0$ is the atomic density, $\hbar$ is the reduced Planck constant, $\omega_{12}$, $\omega_{32}$, $\omega_{31}$ are the transition frequencies of the medium and $d_{12}$, $d_{13}$, $d_{23}$ are the dipole moments of these transitions, $\rho_{11}$, $\rho_{22}$, $\rho_{33}$ are the populations of the 1st, 2nd and 3rd states of the atom respectively, $\rho_{21}$, $\rho_{32}$, $\rho_{31}$ are the non-diagonal elements of the density matrix determining the dynamics of the polarization of the medium, $P$ is the medium polarization. The equations also include $T_{ik}$ relaxation terms. As initial conditions, a half-cycle pulse propagating from left to right was sent from the left boundary of the integration domain $z=0$ to the medium:

$$E(z = 0, t) = E_{01}e^{-\frac{(t-\tau_1)^2}{\tau^2}}, \quad (9)$$

Simultaneously, from the right boundary of the integrating domain $L = 12\lambda_0$, an identical counterpropagating pulse was sent to the medium in the form:

$$E(z = L, t) = E_{02}e^{-\frac{(t-\tau_2)^2}{\tau^2}}, \quad (10)$$

The medium was located in the center of the integration domain between the points $z_1 = 4\lambda_0$ and $z_2 = 8\lambda_0$ The parameters of three-level medium were chosen in a such a manner: $\omega_{12} = 2.69 \cdot 10^{15}$ rad/s ($\lambda_{12} = \lambda_0 = 700$ nm), $\omega_{31} = 1.5\omega_{12}$, $\omega_{32} = \omega_{31} - \omega_{12}$, $d_{12} = 20$ D, $d_{13} = 1.5d_{12}$, $d_{23} = 0$. Atomic density is $N_0 = 10^{14}$cm$^{-3}$. Driving field amplitude $E_0 = 200000$ ESU, duration $\tau = 580$ as, delays $\tau_1 = \tau_2 = 2.5\tau$, relaxation times are $T_{ik}=1$ ns. The system of equations (1)-(10) was integrated numerically. With these parameters, the pulses collide in the center of the medium at the point $z_c = 6\lambda_0$. Then the integrating domain boundaries reflect the pulses back to the medium colliding at the same point, and so on.

## 3. Results of numerical calculations and discussions

Figure 1 shows the spatial-temporal dynamics of the medium polarization, *P*, of the three-level medium. Figures 2-4 show the dynamics of the population difference at each resonance transition.

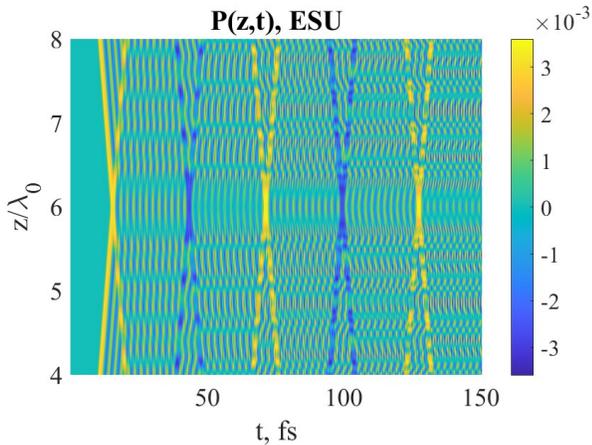

**Fig. 1.** The dynamics of the medium polarization $P(z,t)$ in the three-level medium.

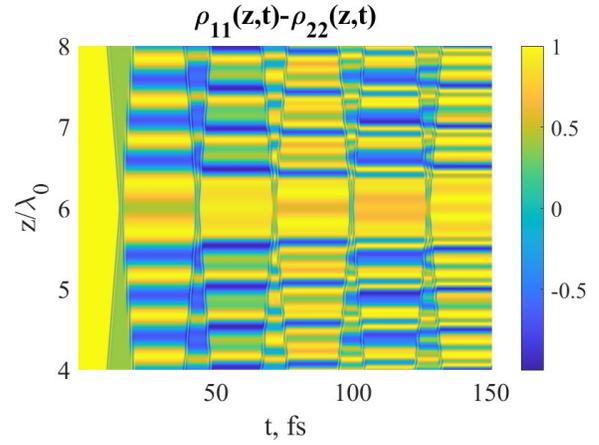

**Fig.2.** Spatial and temporal dynamics of medium population difference $\rho_{11} - \rho_{22}$ in the three-level medium.

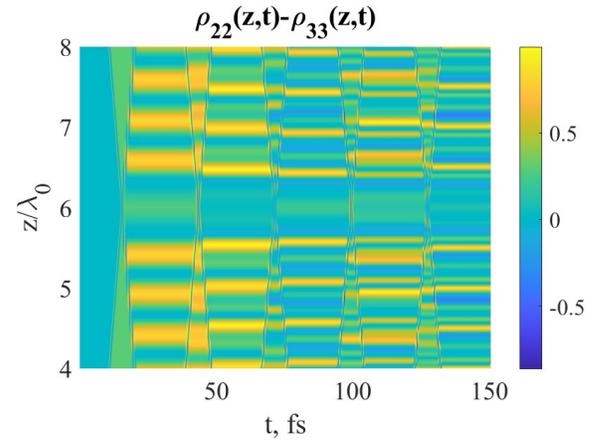

**Fig.3**. Spatial and temporal dynamics of medium population difference $\rho_{22} - \rho_{33}$ in the three-level medium.

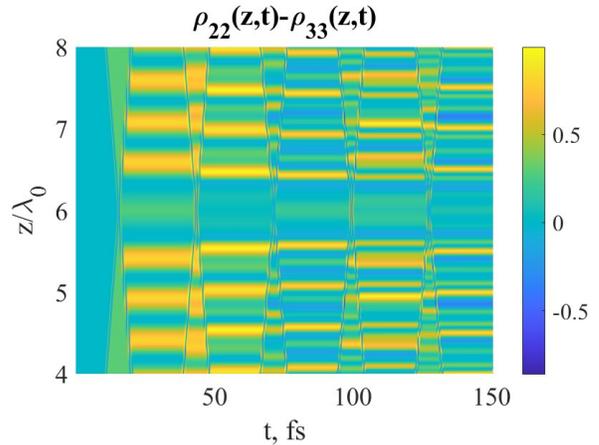

**Fig.4.** Spatial and temporal dynamics of medium population difference $\rho_{11} - \rho_{33}$ in the three-level medium.

It can be seen that between the time intervals 20 and 40 fs in Figures 2-4 a microcavity with Bragg-like mirrors (periodic gratings) in the boundaries is formed after the collision. In addition, in the vicinity of the collision point $z_c = 6\lambda_0$, the inversion of the medium is approximately constant and close to 1 in the transition 1-2, see Fig. 2. As the pulses propagate through the medium prior to the collisions, polarization waves are generated shown in Fig.1. After the collision, these waves interact with each pulse. They are controlled in a coherent manner by the pulse. The result is the formation of a Bragg-type periodic grating. It can be seen that these gratings are controlled after the next collisions. That is, the period and the modulation depth are changed. As shown in Figure 1, complex polarization structures are also formed in the medium.

### 3. Comparison with two-level model

Since many early studies of such systems used the two-level approximation [30-31], it is methodologically interesting to compare the dynamics of gratings in a two-level medium. We have integrated the system of equations similar to (1)-(9) for a two-level medium [30]. The dynamics of the population difference at transition 1-2 is shown in Figure 5.

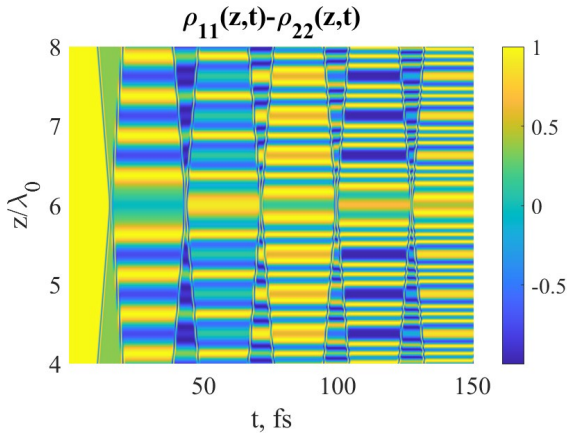

**Fig.5.** Spatial and temporal dynamics of medium population difference $\rho_{11} - \rho_{22}$ in the two-level model. Other parameters are as in Fig.2.

This shows that the inclusion of an additional energy level has not led to the elimination of the effects of grating formation.

### 4. Estimation of the Bragg mirror reflectivity

In order to estimate the reflectivity of the Bragg mirror, we use the standard expression for the reflection coefficient $R(l, \lambda)$ [38] as we have previously used for gratings formed in a two-level medium [34]

$$R(L, \lambda) = \frac{\Omega^2 (\sinh sL)^2}{\Delta k^2 (\sinh sL)^2 + s^2 (\cosh sL)^2}. \quad (11)$$

Where $k = 2\pi n_0/\lambda$ is the propagation constant, $\lambda$ is the incident wavelength, where $\Delta k = k - \pi/\Lambda$ is the detuning $L$ is the grating length, $\Lambda$ is the grating period, $\Omega$ is the coupling coefficient, $\Omega \cong \pi \Delta n_r/\lambda$, $s = \sqrt{\Omega^2 - \Delta k^2}$. The harmonic stationary distribution of the refractive index is assumed according to the expression $n_r = n_0 + \Delta n_r \cdot cos\left(\frac{2\pi z}{\Lambda}\right)$. Assuming $n_0 = 1$, $\Lambda = \lambda_0/2$, $L = 100\lambda_0$, $\Delta n_r = 0.005$ from Eq. (11) we obtain for Bragg wavelength, $\lambda_B = 2\Lambda = \lambda_0 = 700$ nm, $R \simeq 0.7$.

### 5. Conclusions

In this letter, we have theoretically demonstrated a simple way to realize a time-space photonic crystal in a three-level atomic medium. We found that microcavities with Bragg-like mirrors are formed when a train of half-cycle attosecond pulses collides in the medium. This idea opens up a new direction in the study of PTCs and STPCs, providing a novel and convenient way to control the atomic medium with half-cycle pulses. Unlike conventional methods, where Bragg gratings are formed by interfering monochromatic laser beams [38-40], these ideas can be used to create and control Bragg gratings in an ultrafast manner.

We also compared the results of numerical calculations for the dynamics of gratings in two- and three-level media. We found a qualitative similarity in their behavior. This is an indication that the addition of four more levels does not lead to the disappearance of the gratings. This is evidence in favor of the use of a two-level model for problems of coherent half-cycle pulse propagation in a resonant medium.

**Funding.** Russian Science Foundation, project 21-72-10028.

**Acknowledgments.** Author thanks Prof. N.N. Rosanov and Dr. I. Babushkin for helpful discussions.